\documentclass[pre,a4paper,
twocolumn,amsmath,superscriptaddress,showpacs]{revtex4}
\usepackage{graphicx}
\usepackage{bm}
\newcommand{\bee}{\begin{eqnarray}}
\newcommand{\eee}{\end{eqnarray}}
\newcommand{\ba}{\begin{array}}
\newcommand{\ea}{\end{array}}
\newcommand{\aef}{a_{\text{eff}}}

\newcommand{\hs}{^{\text{hs}}}
\newcommand{\bZ}{{\mathbf Z}}
\newcommand{\m}{\mbox{\boldmath $\mu$}}

\usepackage{times}
\usepackage{accents}

\begin{document}
\title{First-order virial expansion of short-time diffusion and sedimentation coefficients of permeable particles suspensions}

\author{Bogdan Cichocki}
 \affiliation{Institute of Theoretical Physics, Faculty of Physics, University of Warsaw, Ho\.za 69,
  00-681 Warsaw, Poland}

\author{Maria L. Ekiel-Je\.zewska}
\email{mekiel@ippt.gov.pl}
 \affiliation{Institute of Fundamental Technological Research,
             Polish Academy of Sciences, Pawi\'nskiego 5B, 02-106 Warsaw, Poland}

\author{Gerhard N\"agele}
 \affiliation{Institute of Complex Systems, ICS-3,  Forschungszentrum J\"ulich,
              D-52425 J\"ulich, Germany}

\author{Eligiusz Wajnryb}
 \affiliation{Institute of Fundamental Technological Research,
             Polish Academy of Sciences, Pawi\'nskiego 5B, 02-106 Warsaw, Poland}

\date{\today}

\begin{abstract}
For suspensions of permeable
particles, the short-time translational and rotational self-diffusion coefficients, and collective diffusion and sedimentation coefficients are evaluated theoretically.
An individual particle is modeled as a uniformly permeable sphere of a given permeability, with the internal solvent flow described by the Debye-Bueche-Brinkman equation. The particles are assumed to interact non-hydrodynamically by their excluded volumes.
The virial expansion of the transport properties in powers of the volume fraction is performed up to the two-particle level. 
The first-order virial coefficients
corresponding to two-body hydrodynamic interactions are evaluated 
with very high accuracy by the series expansion in inverse powers of the inter-particle distance. Results are obtained and discussed for a wide range of the ratio, $x$, of the particle radius to the hydrodynamic screening length inside a permeable sphere. It is shown that for $x\gtrsim 10$, the virial coefficients of the transport properties are well-approximated by the hydrodynamic radius (annulus) model developed by us earlier for the effective viscosity of porous-particle suspensions.

\end{abstract}


\maketitle

\section{Introduction}
One of the theoretical methods to analyze transport properties in suspensions of interacting colloidal particles is the
virial expansion in terms of the particle volume fraction $\phi$. For suspensions of non-permeable hard spheres with stick hydrodynamic
boundary conditions, virial expansion results for short-time properties are known to high numerical precision up to the three-particle level,
i.e., to quadratic order in $\phi$ for diffusion and sedimentation coefficients \cite{CEW:99,Cichocki:02}, and to third order in $\phi$
for the effective viscosity \cite{Cichocki:03}. The concentration range of applicability of these hard-sphere virial expansion results 
in comparison to simulation data has been discussed in \cite{BanchioASD:08}. 
Our knowledge on virial expansion coefficients of colloidal transport properties is less developed 
when suspensions of solvent-permeable particles are considered.
The theoretical description of their dynamics is more complicated since one needs
to account for the solvent flow also inside the particles. Permeable particle systems are frequently encountered
in soft matter science. Prominent examples of practical relevance which are the subject of ongoing research, are
dendrimers \cite{Likos:02,Likos:2010}, microgel particles \cite{Pyett:05,Richtering:08,Coutinho2008}, 
a large variety of core-shell particles with a dry core and
an outer porous layer \cite{Nommensen:01b,Petekidis:04,Zackrisson:05,Adamczyk2004,Malysa}, fractal aggregates \cite{Saarloos:87}, 
and star-like polymers of lower functionality \cite{LikosRichter:98}.

In a series of recent articles \cite{Abade-JCP:10,ACENW-trans-PRE:10,ACENW-visco-JCP:10,ACENW-visco-JPCM:10}, 
we have explored the generic effect of solvent permeability on the short-time transport
using the model of uniformly permeable colloidal spheres with excluded volume interactions. This simple model is specified by two parameters only,
namely the particle volume fraction $\phi = (4\pi/3) n a^3$, where $n$ is the number concentration and $a$ is the particle radius, 
and the ratio $x$ of the particle radius to the hydrodynamic penetration depth inside a permeable sphere. Large (low) values
of $x$ correspond to weakly (strongly) permeable particles. While the model of uniformly permeable hard spheres
ignores a specific intra-particle structure, it is generic in the sense that the hydrodynamic structure of more complex
porous particles can be approximately accounted for in terms of a mean permeability. In our related previous publications,
using a hydrodynamic multipole simulation method of a very high accuracy \cite{CFHWB:94}, encoded in the {\sc hydromultipole} program package
\cite{CEW:99}, we
have calculated the short-time translational diffusion properties \cite{Abade-JCP:10,ACENW-trans-PRE:10}, 
and the high-frequency viscosity \cite{ACENW-visco-JCP:10,ACENW-visco-JPCM:10} of the permeable spheres
model, as functions of $\phi$ and $x$. These results cover the full range of permeabilities, 
with volume fractions extending up to the liquid-solid transition.

While the simulation results are important for the general understanding of permeability effects in concentrated systems,
for practical use in experimental data evaluation and as input in long-time theories, virial expansion results 
based on a rigorous theoretical calculation are still 
strongly on demand. In fact, the knowledge of the leading-order virial coefficients can be a good starting point in deriving approximate expressions
for transport properties, which may be applicable at concentrations much higher than those where the original (truncated) virial expansion result 
is useful. An example in case is provided in our recent derivation \cite{ACENW-visco-JPCM:10} of a
generalized Sait\^o expression for the effective high-frequency viscosity $\eta_\infty$ of permeable spheres,
based on the second-order concentration expansion result, i.e. a Huggins coefficient calculation, of this property.

In \cite{ACENW-visco-JCP:10}, we have performed virial expansion calculations for $\eta_\infty$. 
We have investigated therein a simplifying hydrodynamic radius model (HRM),
where a uniformly permeable sphere of radius $a$ is described by a spherical annulus particle with an inner hydrodynamic radius $\aef(x) < a$,
and unchanged excluded-volume radius $a$. In this annulus model (HRM), the Huggins coefficient describing two-body viscosity contributions
has been evaluated and shown
to be in a remarkably good agreement with the precise numerical data for porous particles, characterized by a wide range of permeabilities realized in experimental systems.

In the present article, the aforementioned theoretical work on the virial expansion of the high-frequency
viscosity is generalized to short-time diffusion properties. In Sec.~\ref{S2}, the virial expansion is performed for 
the translational and rotational self-diffusion coefficients $D_t$ and $D_r$, respectively, the sedimentation coefficient $K$,
and the associated collective diffusion coefficient $D_C=KD_0^t/S(0)$. Here, $S(0)$ is 
the small-wavenumber limit of the static structure factor, and $D_0^t$ is the single-particle translational diffusion coefficient. 
In Sec.~\ref{S3}, we provide highly accurate numerical values for the
first-order (i.e., two-particle) virial coefficients, $\lambda_{t}(x)$, $\lambda_r(x)$, $\lambda_C(x)$ and $\lambda_K(x)$, 
of $D_t$, $D_r$, $D_C$ and $K$, respectively, in the full range of permeabilities. In Sec.~\ref{S4}, we
also recalculate these virial coefficients approximately on the basis of the simplifying annulus model (HRM). In Sec.~\ref{S5}, we conclude that in the range $x \gtrsim 10$ typical of many permeable particle systems, the virial coefficients are well approximated
by the annulus model.

\section{Theory}\label{S2}
We consider a suspension made of a fluid with shear viscosity $\eta_0$ and identical porous particles of radius $a$.
The fluid flow is characterized by Reynolds number Re$<<1$. 
Outside the particles, the fluid velocity ${\bm v}$ and pressure $p$ satisfy the Stokes equations \cite{Kim-Karrila:1991,Happel-Brenner:1986},
\bee
    \eta_0\;\! {\bm \nabla}^2 {\bm v}({\bm r})  - {\bm \nabla} p({\bm r})  &=& 0 \nonumber \\
    {\bm \nabla} \cdot {\bm v}({\bm r}) &=& 0,
\eee
and inside the particles, the Debye-B\"uche-Brinkman (DBB) equations \cite{Brinkman:47d,DebyeBueche:48},
\bee
\label{eq:DBB}
    \eta_0\;\! {\bm \nabla}^2 {\bm v}({\bm r}) - \eta_0\;\! \kappa^2 \left[{\bm v({\bm r})} -
    {\bm u}_i({\bm r}) \right] - {\bm \nabla} p({\bm r})  &=& 0 \nonumber \\
    {\bm \nabla} \cdot {\bm v}({\bm r}) &=& 0,
\eee
where $\kappa^{-1}$ is the hydrodynamic
penetration depth. The skeleton of the particle $i$, centered at ${\bf r}_i$,   moves rigidly with the local  velocity 
${\bf u}_i({\bf r}) = {\bf U}_i + {\bm{\omega}}_i \times \left( {\bf r} - {\bf r}_i \right)$,
determined by the translational and rotational velocities ${\bf U}_i$ and ${\bm{\omega}}_i$ of the particle $i$, respectively. The fluid velocity and  stress tensor are continuous across a particle surface.
The effect of the particle porosity is therefore described by the ratio $x$ of the particle radius $a$ to the hydrodynamic screening length $\kappa^{-1}$ of the porous material inside the particle, i.e.
\bee
x &=& \kappa a.
\eee 

Owing to linearity of the Stokes and DBB equations and the boundary conditions, 
the particle velocity $\mathbf{U}_{i}$ depends linearly on the external forces $\mathbf{F}_{j}$ exerted on a particle $j$. In particular, for two interacting spherical particles, $1,2$, in the absence of external torques and flows,
\begin{eqnarray}
\mathbf{U}_{1} &=&\bm{\mu }_{11}^{tt}({\bf r}_1,{\bf r}_2)\cdot \mathbf{F}_{1}+\bm{%
\mu }_{12}^{tt}({\bf r}_1,{\bf r}_2)\cdot \mathbf{F}_{2} \label{tt1}\\
\mathbf{U}_{2} &=&\bm{\mu }_{21}^{tt}({\bf r}_1,{\bf r}_2)\cdot \mathbf{F}_{1}+\bm{%
\mu }_{22}^{tt}({\bf r}_1,{\bf r}_2)\cdot \mathbf{F}_{2}. \label{tt2}
\end{eqnarray}
In this paper, the two-particle translational-translational mobility matrices $\bm{\mu }_{ij}^{tt}({\bf r}_1,{\bf r}_2)$ are evaluated using the multipole method of solving the Stokes an DBB equations~\cite{CFHWB:94}.
The cluster expansion of the above mobility matrices reads,
\begin{equation}
\bm{\mu }_{ij}^{tt}({\bf r}_1,{\bf r}_2)=\mu _{0}^{t}\delta _{ij}\mathbf{1}+\bm{\mu }%
_{ij}^{tt(2)}(\mathbf{r}),\label{mut2}
\end{equation}
where $\mathbf{r}=\mathbf{r}_{2}-\mathbf{r}_{1}$ and
\begin{equation}
\mu _{0}^{t}=\frac{1}{4\pi \eta_0 A_{10}}\label{mut1}
\end{equation}
is the single porous-particle translational mobility. Here $A_{10}$ is a single porous-particle scattering coefficient~\cite{Cichocki_Felderhof_Schmitz:88}, given explicitly in Appendix \ref{AAA}. For a non-permeable hard sphere with stick boundary conditions, $A_{10}\hs={3a}/{2}$.

The single particle scattering coefficients $A_{l\sigma}$, with $l=1,2,3,4,...$ and $\sigma=1,2,3$ ~\cite{Cichocki_Felderhof_Schmitz:88}, are essential to perform the multipole expansion. 
They determine the corresponding multipoles of the hydrodynamic force density on a particle immersed in an ambient flow; examples are Eqs.~\eqref{mut1} and \eqref{mur1}. The same coefficients $A_{l\sigma}$ specify also the corresponding multipoles of the fluid velocity, reflected (scattered) by a particle 
 immersed in a given ambient flow. This is why $A_{l\sigma}$ are called ``scattering coefficients''. In the multipole approach, differences in the internal structure of particles (e.g. solid, liquid, gas, porous, core-shell, stick-slip), are fully accounted by different scattering coefficients. The other parts of the 
multipole algorithm need not to be changed. The scattering coefficients are the matrix elements of two single-particle friction operators, ${\bf Z}_0$ and $\hat{\bf Z}_0$, which determine the hydrodynamic force density exerted by a given ambient flow on a motionless and a freely moving particle, respectively.

In the multipole expansion method, the two-particle mobility $\bm{\mu }^{(2)}(1,2)$ (e.g. translational one, as in Eq.~\eqref{mut2}, or rotational one, as in Eq.~\eqref{mur2}) 
is expressed in terms of the single-particle friction operators, ${\bf Z}_0(i)$ and $\hat{\bf Z}_0(i)$, with $i=1,2$, and the Green operator ${\bf G}(1,2)$. The later relates the flow outgoing from particle 2 and incoming on particle 1. We can write $\bm{\mu }^{(2)}$  as an infinite scattering series,
\bee
\m^{(2)} &=& \m_0 \bZ_0 (1 + {\bf G}\hat{\bZ}_0)^{-1} {\bf G} 
\bZ_0 \m_0 \nonumber \\
&=& \m_0 \bZ_0 {\bf G}
\bZ_0 \m_0 - \m_0 \bZ_0 {\bf G}\hat{\bZ}_0{\bf G} 
\bZ_0 \m_0 + .... \label{scattmobility}
\eee
Since the multipole matrix elements of the Green tensor {\bf G} scale as  inverse powers of the interparticle distance $r$,  Eq.~\eqref{scattmobility} corresponds to a power series in $1/r$. Truncating the expansion at order $1/r^{1000}$, we obtain a very high precision of the mobility calculation, actually much higher than needed for any practical applications. 

In the present work, we investigate the short-time dynamics, at time scales $t << a^2/D_0^t$, where 
\begin{equation}
D_{0}^{t}=k_{B}T\mu _{0}^{t}\label{self1}
\end{equation}
is the single-particle translational diffusion coefficient, with the Boltzmann constant $k_B$ and temperature $T$. On the short-time scale, the system is described by
the equilibrium particle distribution \cite{hansen}. 
In the further analysis, we will need only the 
small-concentration limit $g_0(r)$ of 
the pair distribution function, where $r$ is the interparticle 
distance. 
For particle-particle direct interactions
described by a pair potential $V(r)$, this pair distribution is  $g_0(r)=\exp(-V(r)/k_{B}T)$. For non-overlapping spheres of radius $a$,
\bee
g_0(r) &=& \left\{ \ba{ll} 0& \mbox{ for }\; r\le 2a,\\ 1&\mbox{ for }\; r > 2a.\ea \right.\label{gg}
\eee
The first-order terms in the virial expansion of the short-time transport coefficients
are obtained by averaging the corresponding two-particle mobility elements. 
As the result, the virial coefficients are obtained as integrals, which involve the mobility elements 
and $g_0(r)$.

The first-order virial expansion of the short-time translational self-diffusion coefficient has the form,
\begin{equation}
D_{t}=D_{0}^{t}(1+\lambda _{t}\phi + {\cal O}(\phi^2)).\label{virtt}
\end{equation}
The coefficient $\lambda _{t}$ is given by the relation~\cite{Cichocki_Felderhof:88},
\begin{equation}
\lambda _{t}=8\int_{1}^{+\infty }g_0(R)J_{t}(R)R^{2}dR,\label{laS}
\end{equation}
with $\mathbf{R=r/}2a$ and
\begin{equation}
J_{t}(R)=\frac{1}{\mu _{0}^{t}}\mbox{Tr}\,\bm{\mu }_{11}^{tt(2)}(\mathbf{r}).
\end{equation}
Here, $\mbox{Tr}$ denotes the trace operation.
For the sedimentation coefficient, one obtains,
\begin{equation}
K=1+\lambda _{K}\phi +... \label{viK}
\end{equation}
where
\begin{eqnarray}
\lambda _{K} &=&\frac{2}{5a^{3}}A_{12}+\frac{8}{a}A_{10}\int_{0}^{+\infty }%
\left[ g_0(R)-1\right] RdR \\
&&+8\int_{1}^{+\infty }g_0(R)J_{K}(R)R^{2}dR,\label{laK}
\end{eqnarray}
and
\begin{equation}
J_{K}(R)=\frac{1}{\mu _{0}^{t}}\mbox{Tr}\left[ \bm{\mu }_{11}^{tt(2)}(\mathbf{r}%
)+\bm{\mu }_{12}^{tt(2)}(\mathbf{r})-\mathbf{T}_{0}(\mathbf{r)}\right].
\end{equation}
In the above expression,
\begin{equation}
\mathbf{T}_{0}(\mathbf{r)=}\frac{\mathbf{1+}\widehat{\mathbf{r}}\widehat{%
\mathbf{r}}}{8\pi \eta_0 r},
\end{equation}
is the Oseen tensor and
$\widehat{\mathbf{r}}=\mathbf{r/}r$.

The scattering coefficient $A_{12}$ for a porous particle~\cite{Cichocki_Felderhof_Schmitz:88} is given explicitly in
Appendix~\ref{AAA}. For a non-permeable hard sphere with the stick boundary conditions, $A_{12}\hs={5a^{3}}/{2}$.

The collective diffusion coefficient is given by
\bee
D_C=D_0^t K/S(0),\label{defDC}
\eee 
where $S(0)$ is the zero-wavenumber limit of the static structure factor, $S(0)=\lim_{q\rightarrow 0}S(q)$.
The first-order virial expansion of Eq. \eqref{defDC} has the form, 
\bee
D_C&=&D_0^t(1+\lambda_C\phi+...).\label{lac}
\eee 
For the non-overlapping spheres~\cite{hansen},
\bee
S(0)=1-8\phi + {\cal O}(\phi^2).
\eee
In this case, 
\bee
\lambda_C=\lambda_K+8.\label{CK}
\eee
The relation \eqref{CK}  follows from Eqs.~\eqref{viK}, \eqref{defDC} and \eqref{lac}.  

We proceed by analyzing the short-time rotational self-diffusion coefficient.
In the absence of external forces and flows, the two-particle rotational-rotational mobility matrices $\bm{\mu }_{ij}^{rr}({\bf r}_1,{\bf r}_2)$ satisfy the relation analogical to Eqs.~\eqref{tt1}-\eqref{tt2}, with the translational velocities replaced by the rotational ones, and the forces replaced by the torques. The two-particle cluster expansion now reads,
\begin{equation}
\bm{\mu }_{ij}^{rr}({\bf r}_1,{\bf r}_2)=\mu _{0}^{r}\delta _{ij}\mathbf{1}+\bm{\mu }%
_{ij}^{rr(2)}(\mathbf{r}),\label{mur2}
\end{equation}
with
\begin{equation}
\mu _{0}^{r}=\frac{1}{8\pi \eta_0 A_{11}}.\label{mur1}
\end{equation}
The scattering coefficient $A_{11}$ for a porous particle~\cite{Cichocki_Felderhof_Schmitz:88} is given in Appendix~\ref{AAA}.
For a non-permeable hard sphere with the stick boundary conditions, ${A_{11}\hs}={a^{3}}$.

The virial expansion of the rotational self-diffusion coefficient is
\begin{equation}
D_{r}=D_{0}^{r}(1+\lambda_{r}\phi +...),
\end{equation}
where
\begin{equation}
D_{0}^{r}=k_{B}T\mu _{0}^{r}\label{Dr1}
\end{equation}
and
\begin{equation}
\lambda_{r}=8\int_{1}^{+\infty }g_0(R)J_{r}(R)R^{2}dR,\label{laR}
\end{equation}
with
\begin{equation}
J_{r}(R)=\frac{1}{\mu _{0}^{r}}\mbox{Tr}\,\bm{\mu }_{11}^{rr(2)}(\mathbf{r}).
\end{equation}


\section{Results}\label{S3}
To evaluate the first order virial coefficients $\lambda\!=\!\lambda_t,\lambda_K,\lambda_r$, we calculate two-particle mobility matrix elements, performing a series expansion in powers of $1/r$ up to the order 1000, as described in 
Sec.~\ref{S2}. Integration with respect to the particle positions has been performed analytically term by term, using the expressions given in the previous section. The virial coefficients $\lambda$ have been evaluated for a wide range of $x$. Selected results are listed in Table~\ref{tab1}. All displayed digits are significant. (More data are available on request.)
\begin{table}[h]
\caption{First virial coefficients $\lambda_{t}$, $\lambda_K$ and $\lambda_r$ for the short-time translational self-diffusion, sedimentation and rotational self-diffusion, respectively.}\label{tab1}
\begin{tabular}{rlll}
\hline
$x$&\hspace{0.4cm}$\lambda_{t}$&\hspace{0.4cm}$\lambda_K$&\hspace{0.4cm}$\lambda_r$\\
\hline
\hline
3 \; \;     &-0.2497\; \;    &   -3.4451\; \;    &  -0.03257  \\
4 \; \;       & -0.4159\;\;   & -4.1066\;\;    & -0.06336\\
5  \; \;    &-0.5692\; \;    &   -4.5539\;\; &     -0.09682\\
6  \; \;      &-0.7021       &   -4.8722      &    -0.12956\\
7  \; \;   &   -0.8149       &   -5.1084      &    -0.16012 \\
8 \; \;    &   -0.9102       &   -5.2898      &    -0.18802 \\
9 \; \;    &   -0.9909       &   -5.4328      &    -0.21327 \\
10 \; \;      &-1.0598        &   -5.5480     &     -0.23606\\
11 \; \;    &  -1.1190        &   -5.6426     &     -0.25662\\
13 \; \;   &   -1.2151        &   -5.7884     &     -0.29208\\
16 \; \;   &   -1.3202        &   -5.9380     &     -0.33426\\
18 \; \;   &        -1.3730  & -6.0095  & -0.35692\\
20 \; \;      &-1.4161        &   -6.0662     &     -0.37628\\
30  \; \;     &-1.5499        &   -6.2335     &     -0.44202\\
40 \; \;      & -1.6190  & -6.3149  & -0.48007\\
50  \; \;     &-1.6610        &   -6.3628     &     -0.50497\\
65 \; \; &     -1.7001   & -6.4064  & -0.52959\\
100 \; \;     &-1.7460        &   -6.4563     &     -0.56075\\
$\infty$\; \; &-1.8315        &   -6.5464     &     -0.63054\\
\hline

\end{tabular}
\end{table}

\begin{figure}[h]
\includegraphics[width=8.5cm]{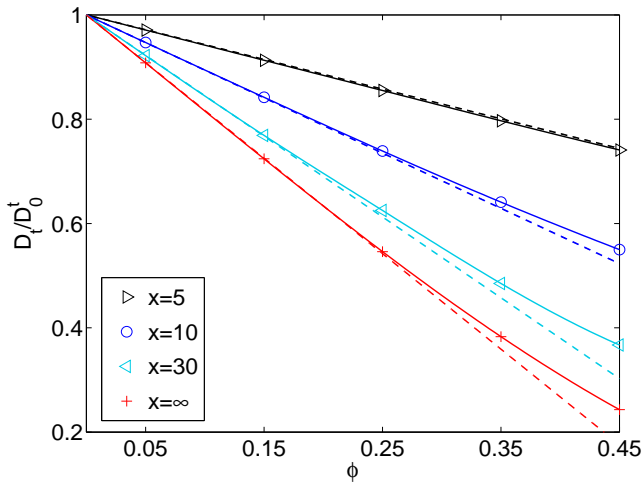}
\caption{Translational self-diffusion coefficient $D_t$ for a suspension of porous particles. Symbols connected by splines (solid lines): accurate simulation results from Ref.~\cite{Abade-JCP:10}. Dashed straight lines: first-order virial expansion calculated in this work.}\label{fig1}
\end{figure}

In Fig.~\ref{fig1}, 
the first order viral expansion of the short-time translational self-diffusion coefficient of a porous-particle suspension is compared with the accurate simulation results performed in  Ref.~\cite{Abade-JCP:10} for volume fractions $\phi \le 0.45$. The first order virial expansion, see Eq. \eqref{virtt}, can be used as an accurate approximation of $D_t$ in a very wide range of volume fractions, even for relatively large values of $x$ (i.e. low permeabilities). 

In contrast, values of the sedimentation coefficient $K$ differ significantly from the first-order virial estimation already at rather small volume fractions, see Ref.~\cite{Abade-JCP:10}. Our values of $\lambda_K(\infty)$ reproduce with a higher accuracy the classic Batchelor's result \cite{BatchelorSedimentation:72} for non-permeable hard spheres. For uniformly porous particles, the boundary collocation method was applied by Chen and Cai \cite{Chen_Cai:99} to evaluate $\lambda_K=-3.46, \;-5.50,\;-6.23,\;-6.44$ for $x^2=10^{\alpha}$, with $\alpha=1,2,3,4$,  respectively.
Comparing their results with our very accurate values, $\lambda_K=-3.5723,\; -5.5480,\; -6.2504,\; -6.4563$, we conclude that the uncertainty of their results is decreasing from 3\% at $\alpha=1$ to 0.3\% at $\alpha=4$.
The accuracy of the boundary collocation method is worse at smaller values of $x$, i.e. for larger permeabilities.

\begin{figure}[h]
\includegraphics[width=8.5cm]{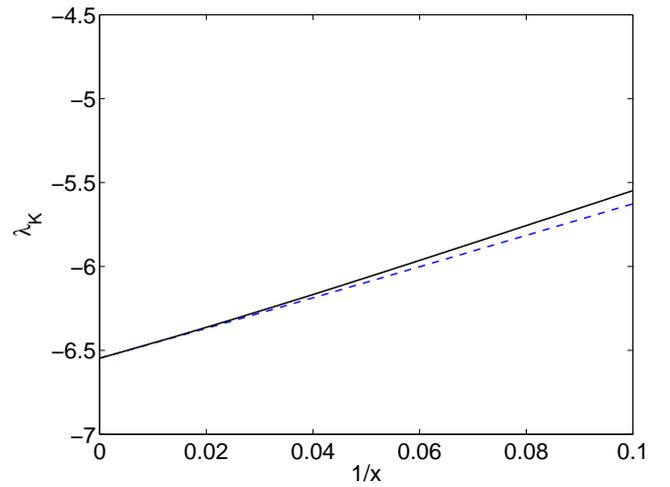}
\caption{Two-particle sedimentation virial coefficient $\lambda_K(x)$. Our precise results for porous particles (solid line) are well-approximated by the annulus model (dashed line).  } 
\label{fig2}
\end{figure}

\begin{figure}[h]
\includegraphics[width=8.5cm]{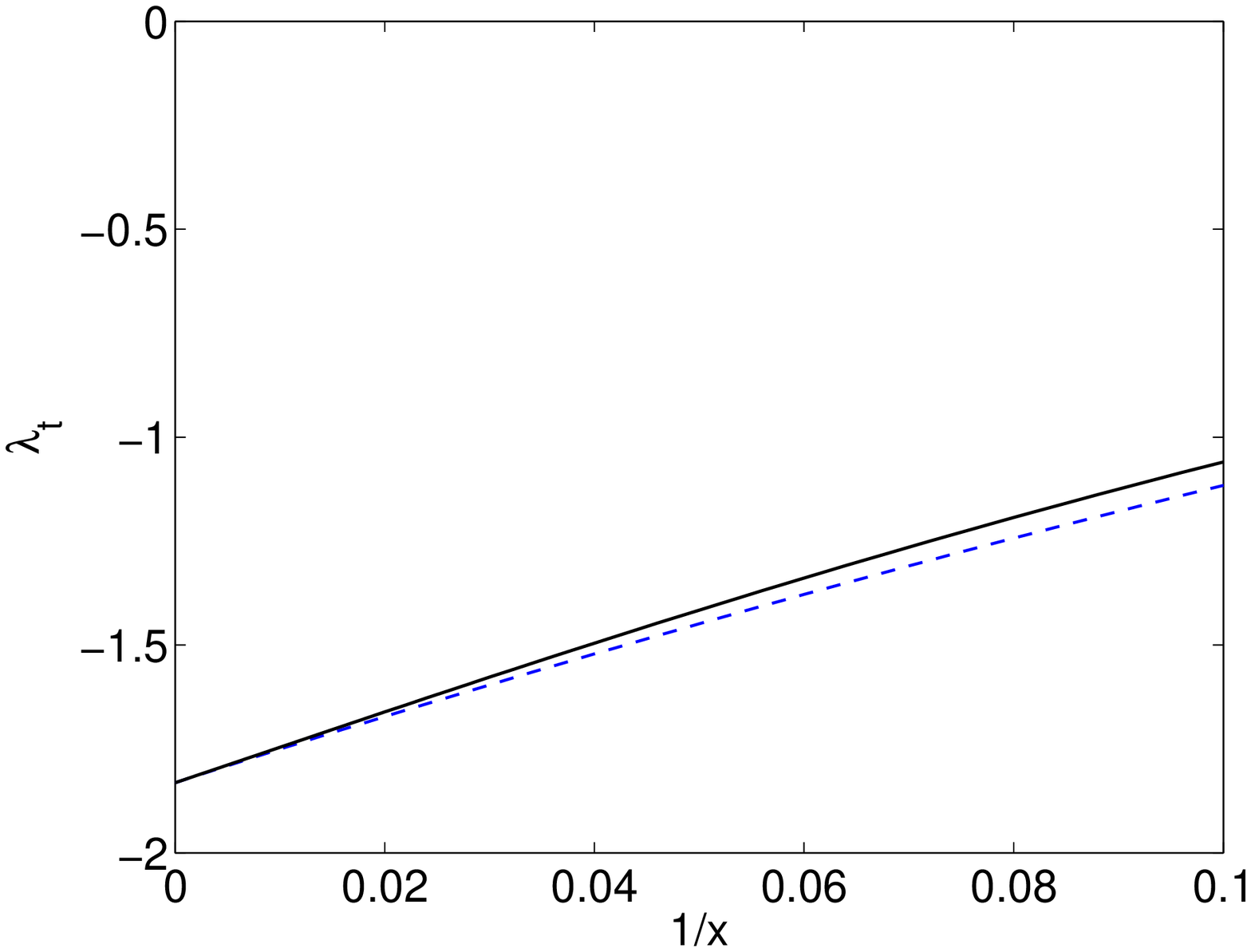}
\caption{Two-particle translational self-diffusion virial coefficient $\lambda_{t}(x)$. Our precise results for porous particles (solid line) are well-approximated by the annulus model (dashed line). }
\label{fig3}
\end{figure}

\begin{figure}[h]
\includegraphics[width=8.5cm]{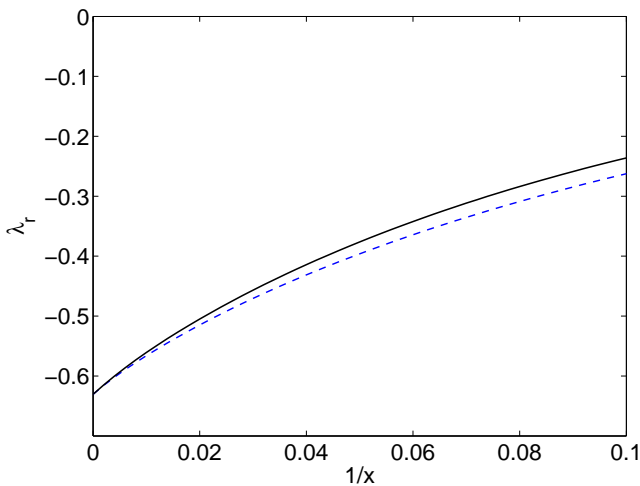}
\caption{Two-particle rotational self-diffusion virial coefficient  $\lambda_r(x)$. Our precise results for porous particles (solid line) are well-approximated by the annulus model (dashed line).}
\label{fig4}
\end{figure}

In Figs.~\ref{fig2}-\ref{fig4}, the first-order virial coefficients $\lambda$ are plotted versus the porosity parameter $1/x$, for $1/x\le 0.1$, i.e. for $x\ge 10$. For $1/x=0$, i.e. for $x=\infty$, the limit of a non-permeable hard sphere with radius $a$ is recovered, $\lambda\hs=\lambda(\infty)$.
As shown in Fig.~\ref{fig2}, for a low permeability, the coefficient $\lambda_K~\approx~\lambda_K\hs~+~10/x$ is approximately a linear function of $1/x$.
The coefficients $\lambda_{t}$ and $\lambda_r$ as functions of $1/x$ are shown in Figs.~\ref{fig3} and~\ref{fig4}.

\newpage
\section{Annulus model}\label{S4}
In the annulus model~\cite{FelderhofCichocki}, a particle suspended in a viscous fluid is characterized by two radii, $a_<$ and $a_>$. Its hydrodynamic interactions are governed by the smaller radius $a_<$. In addition, there exist also direct pair interactions. Two particles cannot come too close to each other, with the no-overlap radius equal to $a_>$. For such a model, the first-order virial expansion of transport coefficients has been performed with respect to the volume fraction $\phi_>=(4\pi/3) n a_>^3$. The corresponding first-order virial coefficients $\lambda^A=\lambda^A_{t},\lambda_K^A,\lambda^A_r$  have been evaluated as functions of $\epsilon$, where
\bee
\epsilon &=&\dfrac{a_>-a_<}{a_<}.\label{defann}
\eee 
The method used to determine $\lambda^A$ is described in Appendix~\ref{A}. The calculated values are listed in Table \ref{tab2}.

\begin{table}[h]
\caption{First-order virial coefficients $\lambda^A_{t}$, $\lambda_K^A$ and $\lambda^A_r$ for the short-time translational self-diffusion, sedimentation and rotational self-diffusion, respectively, for a suspension of the annulus particles.}\label{tab2}
\begin{tabular}{rlll}
\hline
$\epsilon$\; \;\; \;&\hspace{0.4cm}$\lambda^A_{t}$&\hspace{0.4cm}$\lambda_K^A$&\hspace{0.4cm}$\lambda^A_r$\\
\hline
\hline
0.00 \; \; &  -1.8315 \; \;     &  -6.5464  \; \;     & -0.63055\\
0.01 \; \; &  -1.7523 \; \;     &  -6.4601  \; \;     & -0.56666\\
0.02 \; \; &  -1.6793 \; \;     &  -6.3769  \; \;     & -0.51671\\
0.03  \; \;     & -1.6109 \; \;     &  -6.2962  \; \;     & -0.47417\\
0.04 \; \;     &  -1.5466 \; \;     &  -6.2179  \; \;     & -0.43699\\
0.05 \; \;     &  -1.4860 \; \;     &  -6.1419  \; \;     & -0.40402\\
0.06 \; \;     &  -1.4286 \; \;     &  -6.0680  \; \;     & -0.37451\\
0.07 \; \;     &  -1.3743 \; \;     &  -5.9962  \; \;     & -0.34791\\
0.08 \; \;     &  -1.3228 \; \;     &  -5.9263  \; \;     & -0.32381\\
0.09 \; \;     &  -1.2739 \; \;     &  -5.8582  \; \;     & -0.30189\\
0.10 \; \;     &  -1.2274 \; \;     &  -5.7918  \; \;     & -0.28187\\
0.11 \; \;     &  -1.1832 \; \;     &  -5.7272  \; \;     & -0.26354\\
0.13 \; \;     &  -1.1008 \; \;     &  -5.6027  \; \;     & -0.23122\\
0.18 \; \;     &   -0.9253 \; \;     &  -5.3166  \; \;    & -0.16974\\ 
0.24 \; \;     &  -0.7595 \; \;     &  -5.0135  \; \;     & -0.12034\\
0.31 \; \;     &  -0.6111 \; \;     &  -4.7051  \; \;     & -0.08296\\
0.45 \; \;     &  -0.4093 \; \;     &  -4.1986  \; \;     & -0.04242\\
0.66 \; \;     &  -0.2401 \; \;     &  -3.6278  \; \;     & -0.01775\\
\hline
\end{tabular}
\end{table}

Now we are going to compare the first-order virial coefficients, calculated in the previous section for porous particles, with the corresponding results, obtained in this section for the annulus model, called also the hydrodynamic radius model (HRM). A similar comparison has been done in Ref.~\cite{ACENW-visco-JCP:10} for the effective viscosity. The key concept in this procedure is the hydrodynamic radius of a porous particle. 
For the translational diffusion (self-diffusion and sedimentation), the hydrodynamic radius $\aef^{t}$ is obtained from the single-particle translational diffusion coefficient, with the use of the relation,
\bee
D_0^t &=& \dfrac{k_BT}{6\pi\eta_0 \aef^{t}}.
\eee
The dependence of $\aef^t$ on the porosity parameter $x$ follows from Eqs. \eqref{mut1} and \eqref{self1}, which determine the translational diffusion coefficient of a single porous particle \cite{Brinkman:47d,DebyeBueche:48,RFJ:78}, 
and Eq. \eqref{A1}, which specifies the scattering coefficient $A_{10}$. Explicitly,
\bee
{\aef^{t}}(x) &=& a \,\dfrac{2 x^2 (x - \tanh(x))}{ 2x^3 + 3(x - \tanh(x))}. \label{aefK}
\eee

For the rotational self-diffusion, the hydrodynamic radius $\aef^{r}$ is obtained from the rotational diffusion coefficient of a single porous particle \cite{FelderhofDeutch:75,RFJ:78}, with the use of the relation,
\bee
D_0^r &=& \dfrac{k_BT}{8\pi\eta_0 (\aef^{r})^3}.
\eee
The dependence of $\aef^r$ on the porosity parameter $x$ follows from 
 Eqs. \eqref{mur1} and \eqref{Dr1}, with the scattering coefficient $A_{11}$ given by Eq. \eqref{A2}. As the result,
\bee
{\aef^{r}}(x) &=& a\, \left[  1 + \dfrac{3}{x^2} - \dfrac{3 \coth(x)}{x}
\right]^{1/3}.\label{aefR}
\eee
For large $x$, 
\bee
{\aef^{r}}(x) &=& {\aef^{t}}(x) + {\cal O}\left(1/x^2\right).
\eee

\begin{figure}[h]
\includegraphics[width=6.5cm]{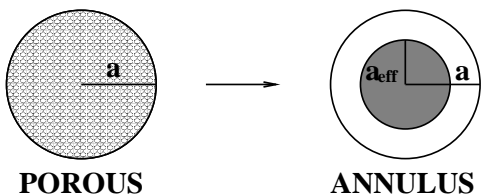}
\caption{The annulus (or hydrodynamic radius) model of a porous particle.}
\label{ann}
\end{figure}

A porous particle of radius $a$ and the porosity parameter $x$  is modeled as an annulus particle, see Fig.~\ref{ann}, by matching 
its geometrical radius $a$ to the annulus no-overlap radius $a_>=a$.
The smaller annulus radius $a_<=\aef$ is determined by the effective hydrodynamic radius $\aef$, given in Eqs. \eqref{aefK}-\eqref{aefR}. 

In this way, the annulus parameter $\epsilon$, defined in Eq. \eqref{defann}, becomes the following function of $x$,
\bee
\epsilon(x) &=&\dfrac{a-\aef(x)}{\aef(x)},\label{defanne}
\eee 
with $\aef(x)$ determined by Eqs.~\eqref{aefK} and \eqref{aefR}.

In Figs.~\ref{fig2}-\ref{fig4}, the annulus coefficients $\lambda^A(\epsilon)$, with $\epsilon(x)$ specified by Eq.~\eqref{defanne} (dashed lines) are compared to the porous-particle virial coefficients $\lambda(x)$ (solid lines). For the sedimentation coefficient,
the annulus model is accurate, with a
half-percent relative accuracy already at $x=20$
and a reasonable 3\% precision at $x=5$. For
the translational self-diffusion, the annulus model is
less accurate, but still it gives only a 2\% error
for $x=20$, and a 5\% error for $x=10$, and a 7\% error for $x=5$.
The least accurate is the annulus prediction for the
rotational self-diffusion, with a 5\% error for
$x=20$ and a 11\% error for $x=10$. Summarizing, the annulus  model (HRM) approximates well the first virial coefficients of porous particles suspensions, in the range of intermediate and small particle permeability (i.e. for moderate and large values of $x$). 
 
\section{Conclusions}\label{S5}

In this paper, the short-time
diffusion properties of dilute
suspensions of uniformly porous spherical particles have been investigated. 
The first-order virial coefficients $\lambda$ of the diffusion and sedimentation coefficients have been evaluated as functions of permeability. Values of $\lambda$ are 
well-approximated by the annulus (hydrodynamic radius)  model, if the parameter $x$
is sufficiently large, i.e. the permeability
is sufficiently low. 
Systematically, the annulus
approximation slightly underestimates the
virial coefficients of porous particles
suspensions. For rotational diffusion, a reasonable 5\% accuracy of this model is reached at $x\gtrsim 20$.
For translational diffusion (collective and self), the comparable or even better 3-5\% precision is obtained already for $x\gtrsim 10$. For the  sedimentation coefficient, the accuracy is even higher (a 3\% precision already at $x=5$), owing to much larger absolute values of $\lambda_K=8+\lambda_C$. 

The annulus model is expected to work well also 
at larger volume fractions, if the porosity parameter $x$ 
is sufficiently large. 
The accuracy of this approximation at larger volume fractions will be investigated in a separate publication. 
Moreover, the annulus model is also useful to estimate diffusion and shear viscosity coefficients of suspensions made of particles characterized by a different internal structure, such as the core-shell. This problem will be the subject of a future study. 

\appendix
\section{Scattering coefficients}\label{AAA}
The scattering coefficients for a uniformly permeable sphere with $x=\kappa a$, where $a$ is the sphere radius and $1/\kappa$ is the hydrodynamic penetration depth, have the form~\cite{RFJ:78},
\begin{eqnarray}
A_{l0} &=& \frac{(2l\!+\!1)g_{l}(x)}{2g_{l-2}(x)} \left [ 1 +
\frac{l(2l\!-\!1)(2l\!+\!1)g_{l}(x)}{(l\!+\!1)x^{2}g_{l-2}(x)} \right
]^{-1}\!a^{2l-1},\nonumber \\ \label{A1}\\
A_{l1} &=& \frac{g_{l+1}(x)}{g_{l-1}(x)}a^{2l+1}, \label{A2} \\
A_{l2} &=& \left [ \frac{2l\!+\!3}{2l\!-\!1}+ \frac{2(2l\!+\!1)(2l\!+\!3)}{(l\!+\!1)x^{2}} \right ]
a^{2}A_{l0}-\frac{2l\!+\!3}{2l\!-\!1}a^{2l+1},\nonumber \\ \label{A3} \\
B_{l2} &=& \left [ 1 + \frac{2(2l-1)(2l\!+\!1)}{(l+1)x^{2}} \right ]a^{2}A_{l2}-a^{2l+3},
\end{eqnarray}
where $l=1,2,...$  and $g_{l}(x) = \sqrt{\pi/2x}I_{l+1/2}(x)$ is the modified spherical Bessel function of the first kind.

\section{The annulus (hydrodynamic radius) model}\label{A}

For a suspension of particles described by the annulus (hydrodynamic radius) model, the virial coefficients %
$\lambda^A$ are functions of the model parameter $\epsilon$, defined by Eq.~\eqref{defann} and listed in Table \ref{tab2}. In this Appendix, we explain how these values have been evaluated.

The first-order virial expansion of $D_t/D_0^t$, $D_C/D_0^t$, $K$ and $D_r/D_0^r$, can be performed with the use of $\phi_<=(4\pi/3)na_<^3$, or $\phi_>$ defined by the analogical expression,
\bee
1+\lambda^A(\epsilon)\phi_> + {\cal O}\left(\phi_{>}^2\right) &=& 1+\bar{\lambda}^A(\epsilon)\phi_< + {\cal O}\left(\phi_{<}^2\right).\hspace{0.9cm}
\eee
Therefore,
\bee
\lambda^A (\epsilon)&=&  
 \dfrac{\bar{\lambda}^A(\epsilon)}{(1+\epsilon)^3}.
\eee
To evaluate $\bar{\lambda}^A(\epsilon)$, we now introduce the dimensionless interparticle distance as $R=r/2a_<$, 
and we replace the pair distribution function from Eq.~\eqref{gg} by the following expression,
\bee
g_0(R) &=& \left\{ \ba{ll} 0& \mbox{ for }\; R\le 1+\epsilon,\\ 1&\mbox{ for }\; R > 1+\epsilon.\ea \right.,\label{ggplus}
\eee
which corresponds to the no-overlap condition at a larger radius $a_>$. Then, we apply the Eqs.~\eqref{laS}, \eqref{laK} and \eqref{laR}, taken  in the non-permeable hard-sphere limit, $x=\infty$. 
We obtain the following expressions,
\bee
\bar{\lambda}^A_{t}(\epsilon )\!\!&=&\!\! 8\int_{1+\epsilon }^{+\infty
}J_{t}(R)R^{2}dR\nonumber\\
&=&\lambda _{t}^{\text{hs}}-8\int_{1}^{1+\epsilon }J_{t}(R)R^{2}dR,
\\
\bar{\lambda}^A_{K}(\epsilon )\!\! &=&\!\!\frac{2}{5a_<^{3}}A_{12}^{\text{hs}}-\frac{4}{a_<}%
A_{10}^{\text{hs}}(1+\epsilon )^{2}\nonumber\\
&+&8\int_{1+\epsilon }^{+\infty }J_{K}(R)R^{2}dR \nonumber \\
&=&\lambda _{K}^{\text{hs}}-8\int_{1}^{1+\epsilon }\bar{J}_{K}(R)R^{2}dR,
\eee\bee
\bar{\lambda}^A_{r}(\epsilon )\!\!&=& \!\!8\int_{1+\epsilon }^{+\infty
}J_{r}(R)R^{2}dR\nonumber\\
&=&\lambda _{r}^{\text{hs}}-8\int_{1}^{1+\epsilon }J_{r}(R)R^{2}dR,
\eee
where
\begin{equation}
\bar{J}_{K}(R)=\frac{1}{\mu _{0}^{t}}\mbox{Tr}\left[ \bm{\mu }%
_{11}^{tt(2)}(\mathbf{R})+\bm{\mu }_{12}^{tt(2)}(\mathbf{R})\right].
\end{equation}
In this Appendix, all the mobility coefficients, the associated functions $J$, and the superscript $^{\text{hs}}$, refer to the hard spheres with the stick boundary conditions.

The above formulas have been applied to compute the functions ${\lambda}^A_{t}(\epsilon )$, ${\lambda}^A_{K}(\epsilon )$ and ${\lambda}^A_{r}(\epsilon )$, listed in Table~\ref{tab2}. The functions  $\bar{\lambda}^A_{t}(\epsilon )$, $\bar{\lambda}^A_{K}(\epsilon )$ and $\bar{\lambda}^A_{r}(\epsilon )$ have been also evaluated and listed in Ref.~\cite{Cichocki:2010}, where they have been applied to model diffusion and rheology of particles with a hard solid core and a thin porous shell.

\end{document}